\documentclass[aps,preprint]{revtex4}%
\usepackage{amsfonts}
\usepackage{amsmath}
\usepackage{amssymb}
\usepackage{graphicx}%
\providecommand{\U}[1]{\protect\rule{.1in}{.1in}}

\begin{document}
\title[ ]{Symmetries and Thermal Radiation: A Classical Derivation of the Planck Spectrum}
\author{Timothy H. Boyer}
\affiliation{Department of Physics, City College of the City University of New York, New
York, New York 10031}
\keywords{}
\pacs{}

\begin{abstract}
A derivation of the Planck spectrum for thermal radiation is given based upon
wave fluctuations within relativistic classical physics. \ The derivation
depends crucially on thermal fluctuations existing above the fundamental
inertial-frame-independent fluctuations of classical zero-point radiation.
\ Such frame-independent zero-point fluctuations exist only in a relativistic
wave theory and cannot exist in a nonrelativistic wave theory. \ Thus such a
classical derivation of the Planck spectrum exists in a Lorentz-covariant
classical theory, such as classical electrodynamics, but not in a
Galilean-covariant theory where all waves are based upon material media.
\ Classical zero-point radiation provides a purely classical alternative to
quanta in the analysis of the Planck spectrum. \ 

\end{abstract}
\maketitle

\section{Introduction}

\subsection{Classical Insight for Blackbody Radiation}

Classical physics sometimes gives insights into physical phenomena which are
hidden in quantum physics. \ Thus when classical explanations are possible,
they are often welcomed. \ Here we show that a purely classical \textit{wave}
explanation for the Planck spectrum of thermal radiation is indeed possible,
but only if we include the idea of classical zero-point radiation, which
zero-point radiation exists only in a \textit{relativistic} classical theory. \ 

Relativistic classical waves, such as electromagnetic waves, propagate through
\textit{vacuum} with the universal speed $c$. \ Such relativistic waves are
completely different from waves in nonrelativistic physics which always travel
through some \textit{material medium} having a mass density and an associated
strain parameter; these parameters determine the wave speed $v$ which speed
has no upper bound. \ Both the \textit{wave} nature of the system and the
\textit{relativistic} covariance are crucial to the analysis leading to the
Planck spectrum within purely classical physics. \ 

\subsection{The General Solution of Maxwell's Equations}

The Planck spectrum for blackbody radiation involves fluctuating
electromagnetic waves which, according to classical theory, satisfy Maxwell's
equations. \ Maxwell's equation are linear in the electric and magnetic fields
with sources in charge and current densities. \ The general solution of a set
of linear differential equations requires both a particular solution and a
solution of the homogeneous (source-free) equations. \ Thus, a general
solution of Maxwell's equations requires both an integral over the specific
charge density and current density, and also a source-free solution of the
equations. \ It is often assumed that giving the particular solution using the
retarded version from the sources is sufficient, and that the source-free
solution \textquotedblleft obviously vanishes.\textquotedblright%
\cite{Lorentz}\cite{Griffiths} \ However, the determination of the source-free
solution of Maxwell's equations within \textit{purely classical physics} is
actually an \textit{experimental} question. \ 

Many physicists have always seen classical electrodynamics without the
homogeneous solution of Maxwell's equation. \ Indeed,
Griffiths\cite{Griffiths} mentions (but only in a footnote) that the choice of
the homogeneous solution that must be made \textit{within purely classical
electrodynamics. \ }Jackson\cite{Jackson1} notes the requirement of the
homogeneous term for the scalar wave equation in connection with the Green
function for the wave operator, but he omits the homogeneous term when writing
the solution of Maxwell's equations. \ Indeed, there is no current textbook of
classical electrodynamics which emphasizes that purely classical
electrodynamics necessitates the choice of a homogenous solution when solving
Maxwell's equations. \ Therefore many instructors assume incorrectly that any
additional homogeneous term must involve a solution \textit{outside of purely
classical physics}. \ This idea is completely false. \ Many physicists forget
that a constant electric field or a plane wave are examples of solutions of
the source-free Maxwell equations. \ Therefore the use of a source free
solution such as a plane wave or a superposition of plane waves does not take
us outside classical electrodynamics. The choice of the homogeneous solution
of Maxwell's equations is an integral part of purely classical theory. As
Coleman states,\cite{Coleman} \textquotedblleft These boundary conditions are
as much a part of the postulates of the theory as the form of the Lagrangian
or the value of the electron charge.\textquotedblright\ \ 

It turns out that for a number of phenomena where Planck's constant $\hbar$
appears, the best choice to explain physical phenomena within
\textit{classical} physics involves random classical radiation with a
Lorentz-invariant spectrum with a \textit{scale constant} chosen as Planck's
constant $\hbar$. \ Such random \textit{classical} radiation is termed
\textquotedblleft\textit{classical} electromagnetic zero-point
radiation.\textquotedblright\cite{Any} \ There is no assumption whatsoever of
\textquotedblleft quantum\textquotedblright\ aspects. \ Thus in classical
physics, Planck's constant $\hbar$ enters the theory as the scale of classical
zero-point radiation. \ Indeed, classical zero-point radiation can be regarded
as a classical alternative to quanta. \ \ In this article, we point out that
the presence of \textit{classical} zero-point radiation allows a natural
classical explanation based upon wave fluctuations for the Planck spectrum for
relativistic wave systems, such as electromagnetic radiation. \ 

\subsection{Thermal Equilibrium for Massive Particles versus Relativistic
Waves}

Thermal equilibrium always involves a \textit{finite density} of
\textit{thermal} energy. \ Thermal equilibrium for massive particles always
deals with a finite number of particles and a finite amount of thermal energy
in a unit volume. \ However, for relativistic waves traveling at speed $c$,
there are always an \textit{infinite} number of waves of all possible
frequencies in a unit volume. \ When the finite density of thermal energy is
shared by instantaneous point-particle collisions, the mechanical potential
function is of no relevance; the potential function enters only between
collisions. \ There is a sharp contrast between sharing the finite thermal
energy among a finite number of particles, compared to sharing a finite
thermal energy among a divergent number of radiation normal modes where the
interaction of the radiation and the matter depends crucially on the
oscillation frequencies of charges. \ This sharp contrast between massive
particle systems and relativistic wave systems suggests that the energy
sharing leading to equipartition, which is suitable for describing massive
particles, may be completely inappropriate for describing thermal equilibrium
for relativistic waves.

The divergent number of relativistic waves in a unit volume means that there
must be some criterion for a distinction among the radiation modes which
assigns more of the \textit{thermal} energy to lower frequency modes and less
\textit{thermal} energy to high frequency modes. \ One such criterion is
Planck's energy-quantum hypothesis assigning energy in quantum units of
$\hbar\omega$. \ The hypothesis of classical zero-point radiation provides an
alternative to quanta which has the advantage of keeping the analysis of
thermal equilibrium purely within classical theory. \ In this article, we show
that indeed the hypothesis of classical electromagnetic zero-point radiation
leads to a natural classical derivation of the Planck spectrum including
zero-point radiation.

\subsection{Outline of the Derivation}

The first part of the article introduces a two-dimensional (one space and one
time) relativistic wave system, and discusses some ideas of classical
zero-point radiation. \ Since many physicists are not experts on the ideas of
relativity, we try to explain the puzzling notion of a spectrum of random
classical radiation (zero-point radiation) which has the same appearance in
every inertial frame. \ The second part of the article treats wave
fluctuations connected to thermodynamic entropy, and shows how the failure to
include zero-point radiation leads to the divergent Rayleigh-Jeans spectrum
with its \textquotedblleft ultraviolet catastrophe,\textquotedblright\ whereas
the inclusion of classical zero-point radiation leads to the Planck spectrum.
\ The third and last part of the article includes various comments on thermal
radiation and the connections between classical and quantum theories. \ 

\section{Massless Waves in Two-Dimensional Spacetime}

\subsection{Relativity in Two-Dimensional Spacetime}

In the following analysis, we treat special relativity for a relativistic
(massless) wave traveling at speed $c$ in two spacetime dimensions. \ We use
this simple \textit{scalar}-wave system since it shows the essential aspects
of thermal equilibrium without the complications of the polarizations of a
\textit{vector} field theory such as classical electrodynamics. \ The
simplification here is like treating the thermodynamics of a nonrelativistic
ideal gas in terms of massive point particles versus treating an ideal gas
using rotating massive spheres of finite radius with an associated rotation axis.

In this two-dimensional relativistic theory, the two spacetime dimensions are
taken as the time $t$ together with one space coordinate $x$.\cite{Zee430}%
\cite{B-84} \ The spacetime satisfies the symmetries of the Lorentz group with
the familiar Lorentz transformations connecting time and space. \ The
space-rotation group is absent because it requires several spatial dimensions. \ 

\subsection{Scalar Waves}

In nonrelativistic physics, the Hamiltonian for a scalar wave $\phi(t,x)~$on a
massive string is given by \ \
\begin{equation}
H=\frac{1}{2}\int dx\left[  \mu\left(  \frac{\partial\phi}{\partial t}\right)
^{2}+Y\left(  \frac{\partial\phi}{\partial x}\right)  ^{2}\right]
\end{equation}
where $\mu$ is the mass per unit length of the medium and $Y$ is the tension.
\ The speed of wave propagation depends upon the parameters of the medium and
given by $v=\sqrt{Y/\mu}$. \ As the mass per unit length $\mu$ becomes ever
smaller, the wave speed $v$ increases without limit in nonrelativistic
physics. \ 

Relativistic physics can be used to describe waves in a material medium or in
vacuum; waves in vacuum corresponds to relativistic (massless) waves which
travel with the speed of light $c$. \ All \ wave speeds $v$ are limited by the
speed $c$ in relativistic physics. \ A relativistic massless wave $\phi\left(
t,x\right)  $ in two-dimensional relativistic spacetime has a Hamiltonian,
\begin{equation}
H=\frac{1}{2}\int dx\left[  \frac{1}{c^{2}}\left(  \frac{\partial\phi
}{\partial t}\right)  ^{2}+\left(  \frac{\partial\phi}{\partial x}\right)
^{2}\right]  \label{HH}%
\end{equation}
where the integral is over all space. \ A harmonic massless scalar wave of
(angular) frequency $\omega$ travels with the speed of light $c$ as
\begin{equation}
\phi_{k}\left(  t,x\right)  =A\cos\left[  kx-\omega t+\theta\left(  k\right)
\right]  , \label{phitx}%
\end{equation}
where $k=\pm\omega/c$ (with the sign depending on the propagation direction of
the wave), $A$ is an amplitude, and $\theta\left(  k\right)  $ is a phase. \ 

\subsection{Isotropic Distribution of Waves}

A single wave would not provide an \textquotedblleft
isotropic\textquotedblright\ distribution since there is a preferred
direction, namely the propagation direction of the wave. \ However, a
\textquotedblleft distribution of waves\textquotedblright\ (set of waves) can
be made \textquotedblleft isotropic\textquotedblright\ (having the same
appearance in every direction) provided for each wave $\phi_{k}\left(
t,x\right)  $, there is another wave $\phi_{-k}\left(  t,x\right)  $ of equal
amplitude but propagating in the opposite direction. \ Two waves traveling in
opposite directions with equal amplitude create a standing wave which is
moving neither to the left nor to the right. \ If there are many paired waves
in the distribution, then there will be many standing waves interfering with
each other, but still providing an isotropic distribution of waves which has
no preferred direction to the left or to the right. \ 

\subsection{Random Radiation}

The paired waves of different frequencies need not be in phase in time.
\ Indeed, classical thermal radiation has always been treated as
\textquotedblleft random radiation\textquotedblright\ ever since the early
work of Planck.\cite{Planck}\ \ Random classical radiation corresponds to
amplitudes and phases of the classical waves satisfying a stochastic process.
\ Einstein and Hopf\cite{EH1910a} showed that for waves spaced closely in
frequency, the randomness could be assigned to the wave phase alone. \ For
individual wave normal modes, the full stochastic description is
employed.\cite{Rice}

\subsection{Lorentz Transformation of the Waves}

Under a Lorentz transformation to a new primed inertial frame,
\begin{equation}
t^{\prime}=\gamma(t-Vx/c^{2})\text{ \ \ \ and \ \ }~x^{\prime}=\gamma\left(
x-Vt\right)  , \label{Ltrans}%
\end{equation}
a harmonic massless wave is still a harmonic wave. \ Thus, the speed
$u^{\prime}$ of the wave in the primed inertial frame is given by the
relativistic velocity addition formula%
\begin{equation}
u^{\prime}=\frac{\pm c-V}{1-\left(  \pm cV/c^{2}\right)  }=\pm c,
\end{equation}
and so remains invariant at $\pm c$. \ The amplitude $A$ of the scalar wave is
unchanged,
\begin{equation}
\phi_{k^{\prime}}^{\prime}\left(  t^{\prime},x^{\prime}\right)  =A\cos\left[
k^{\prime}x^{\prime}-\omega^{\prime}t^{\prime}+\theta\left(  k\right)
\right]  ,
\end{equation}
but now there is a new frequency\cite{Jackson521}
\begin{equation}
\omega^{\prime}=\gamma\left(  \omega-Vk\right)  =\sqrt{\frac{1\mp V/c}{1\pm
V/c}}\omega, \label{omp}%
\end{equation}
and a new wave number
\begin{equation}
k^{\prime}=\gamma\left(  k-V\omega/c^{2}\right)  =\pm\sqrt{\frac{1\mp
V/c}{1\pm V/c}}\frac{\omega}{c} \label{kp}%
\end{equation}
with the $\pm$sign depending upon whether $k=\omega/c$ or $k=-\omega/c$. \ The
phase of the wave in the new primed inertial frame is the same as the phase in
the unprimed inertial frame since, from our Lorentz transformations in Eqs.
(\ref{Ltrans}), (\ref{omp}) and (\ref{kp}), it follows that
\begin{equation}
kx-\omega t=k^{\prime}x^{\prime}-\omega^{\prime}t^{\prime}. \label{pha}%
\end{equation}
\ 

\subsection{A Superposition of Plane Waves}

Waves allow superposition. \ We can imagine a massless wave in a very large
box from $x=-a~$to $x=a$ with periodic boundary conditions. \ The full wave
can be written as a superposition of harmonic waves with different wave
numbers $k_{n}$ so that the total wave is given by%
\begin{equation}
\phi\left(  t,x\right)  =\sum\nolimits_{n=-\infty}^{n=\infty}\,A_{n}%
\cos\left[  k_{n}x-\omega_{n}t+\theta_{n}\right]  , \label{phisum}%
\end{equation}
where $A_{n}$ is the frequency-dependent and direction-dependent amplitude for
the harmonic wave at wave number $k_{n}=2\pi n/\left(  2a\right)  $, and
$\theta_{n}$ is the phase of the associated wave. \ If the phase $\theta_{n}$
is chosen as a random variable distributed uniformly over the interval
$(0,2\pi]$ and distributed independently for each $k_{n}$, then we have random
classical radiation. \ The 3-dimensional analogue of this model was used for
electromagnetic thermal radiation by Planck\cite{Planck} and by Einstein and
Hopf\cite{EH1910a} at the turn of the 20th century.

\subsection{Example of the Correlation Function for Two Waves}

The random radiation in Eq. (\ref{phisum}) involves fluctuations in space and
in time. \ However, we want some stable measure of the spectrum of the
fluctuations, and, for this, we go to the correlation function which involves
an average over the fluctuatig wave. \ A basic understanding of the
correlation function can be gained by considering the familiar example of the
interference of two waves $\phi_{1}(x,t)=A\cos\left[  k_{1}x-\omega
_{1}t+\theta_{1}\right]  $ and $\phi_{2}\left(  x,t\right)  =A\cos\left[
k_{2}x-\omega_{2}t+\theta_{2}\right]  $ of different wave numbers $k_{1}$ and
$k_{2}$, of different frequencies $\omega_{1}=vk_{1}$, $\omega_{2}=vk_{2}$,
and of different phases $\theta_{1}$ and $\theta_{2}$, where $v$ is the wave
velocity. \ The superposition $\phi=\phi_{1}+\phi_{2}$ of the two waves can be
written as
\begin{align}
\phi\left(  x,t\right)   &  =A\cos\left[  k_{1}x-\omega_{1}t+\theta
_{1}\right]  +A\cos\left[  k_{2}x-\omega_{2}t+\theta_{2}\right] \nonumber\\
&  =2A\cos\left[  \left(  \frac{k_{1}+k_{2}}{2}\right)  x-(\frac{\omega
_{1}+\omega_{2}}{2})t+(\frac{\theta_{1}+\theta_{2}}{2})\right] \nonumber\\
&  \times\cos\left[  \left(  \frac{k_{1}-k_{2}}{2}\right)  x-(\frac{\omega
_{1}-\omega_{2}}{2})t+(\frac{\theta_{1}-\theta_{2}}{2})\right]  .
\end{align}
Clearly, the interference of the waves of different wave numbers and of
different frequencies leads to a fluctuating wave, with beats both in space
and in time. \ 

In order to obtain a measure of the spectrum of the fluctuations, we take the
correlation function of the superposition wave at different spatial points $x$
and $x+\Delta x,$ and at different times $t$ and $t+\Delta t$,
\begin{equation}
\left\langle \phi(x,t)\phi\left(  x+\Delta x,t+\Delta t\right)  \right\rangle
=\left\langle \left\{  \phi_{1}\left(  x,t\right)  +\phi_{2}\left(
x,t\right)  \right\}  \left\{  \phi_{1}\left(  x+\Delta x,t+\Delta t\right)
+\phi_{2}\left(  x+\Delta x,t+\Delta t\right)  \right\}  \right\rangle .
\end{equation}
We may average over space or over time or over the phases $\theta$ since any
one of these averages will give the same result. \ Clearly, the correlation
function involves four terms of the basic form $\left\langle \phi_{1}\left(
x+\Delta x,t+\Delta t\right)  \phi_{1}\left(  x+\Delta x,t+\Delta t\right)
\right\rangle $ where the same wave $\phi_{1}$ is considered twice, or of the
basic form $\left\langle \phi_{1}\left(  x,t\right)  \phi_{2}\left(  x+\Delta
x,t+\Delta t\right)  \right\rangle $ where $\phi_{1}$ is considered once and
$\phi_{2}$ is considered once.

If the same wave $\phi_{1}$ or $\phi_{2}$ is considered twice, we use
$\cos\left[  b\right]  =\cos\left[  a\right]  \cos\left[  b-a\right]
-\sin\left[  a\right]  \sin\left[  b-a\right]  $ to obtain
\begin{align}
&  \left\langle \phi_{1}\left(  x+\Delta x,t+\Delta t\right)  \phi_{1}\left(
x+\Delta x,t+\Delta t\right)  \right\rangle \nonumber\\
&  =\left\langle A\cos\left[  k_{1}x-\omega_{1}t+\theta_{1}\right]
A\cos\left[  k_{1}\left(  x+\Delta x\right)  -\omega_{1}\left(  t+\Delta
t\right)  +\theta_{1}\right]  \right\rangle \nonumber\\
&  =\left\langle A^{2}\cos^{2}\left[  k_{1}x-\omega_{1}t+\theta_{1}\right]
\cos\left[  c\right]  \right. \nonumber\\
&  -\left.  A^{2}\cos\left[  k_{1}x-\omega_{1}t+\theta_{1}\right]  \sin\left[
k_{1}x-\omega_{1}t+\theta_{1}\right]  \sin\left[  k_{1}\Delta x-\omega
_{1}\Delta t\right]  \right\rangle \nonumber\\
&  =(A^{2}/2)\cos\left[  k_{1}\Delta x-\omega_{1}\left(  \Delta t\right)
\right]  .
\end{align}
In this equation, it is clear that we find exactly the same result whether we
average over space or over time or over the phase $\theta_{1}$. \ On the other
hand, if the term involves the product of wave $\phi_{1}\left(  x,t\right)  $
with wave $\phi_{2}\left(  x,t\right)  $, the analogous average over space or
over time or over the different phases $\theta_{1}$ and $\theta_{2}$ involves
different arguments because of the differing wave numbers, the differing
frequencies and the different phases, and so gives a vanishing result.
\ Accordingly, the correlation function for the superposition of the two waves
is given by%
\begin{align}
&  \left\langle \phi(x,t)\phi\left(  x,t\right)  \right\rangle \nonumber\\
&  =(A^{2}/2)\cos\left[  k_{1}\Delta x-\omega_{1}\Delta t\right]
+(A^{2}/2)\cos\left[  k_{2}\Delta x-\omega_{2}\Delta t\right]  .
\end{align}
We see that the correlation function has picked out the original plane waves
at $k_{1}$, $\omega_{1},$ and $k_{2}$, $\omega_{2}$. \ 

\subsection{Correlation Function for Random Radiation}

The correlation function for the random radiation in Eq. (\ref{phisum}) can be
obtained by an average over the random phases,
\begin{align}
&  \left\langle \phi\left(  t_{1},x_{1}\right)  \phi\left(  t_{2}%
,x_{2}\right)  \right\rangle \nonumber\\
&  =\left\langle \sum\nolimits_{n=-\infty}^{n=\infty}\,A_{n}\cos\left[
k_{n}x_{1}-\omega_{n}t_{1}+\theta_{n}\right]  \sum\nolimits_{n^{\prime
}=-\infty}^{n^{\prime}=\infty}\,A_{n^{\prime}}\cos\left[  k_{n^{\prime}}%
x_{2}-\omega_{n^{\prime}}t_{2}+\theta_{n^{\prime}}\right]  \right\rangle
_{\theta_{n},\theta_{n^{\prime}}}\nonumber\\
&  =\sum\nolimits_{n=-\infty}^{n=\infty}\sum\nolimits_{n^{\prime}=-\infty
}^{n^{\prime}=\infty}A_{n}A_{n^{\prime}}\left\langle \cos\left[  k_{n}%
x_{1}-\omega_{n}t_{1}+\theta_{n}\right]  \cos\left[  k_{n^{\prime}}%
x_{2}-\omega_{n^{\prime}}t_{2}+\theta_{n^{\prime}}\right]  \right\rangle
_{\theta_{n},\theta_{n^{\prime}}}\nonumber\\
&  =\frac{1}{2}\sum\nolimits_{n=-\infty}^{n=\infty}\,A_{n}^{2}\cos\left[
k_{n}\left(  x_{1}-x_{2}\right)  -\omega_{n}\left(  t_{1}-t_{2}\right)
\right]  \label{pxtpxt}%
\end{align}
where we have averaged over the random phases using $a_{1}=k_{n}x_{1}%
-\omega_{n}t_{1}$, $a_{2}=k_{n^{\prime}}x_{2}-\omega_{n^{\prime}}t_{2}$
\begin{align}
&  \left\langle \cos\left[  a_{1}+\theta_{n}\right]  \cos\left[  a_{2}%
+\theta_{n^{\prime}}\right]  \right\rangle _{\theta_{n},\theta_{n^{\prime}}%
}\nonumber\\
&  =\left\langle \left[  \cos a_{1}\cos\theta_{n}-\sin a_{1}\sin\theta
_{n}\right]  \left[  \cos a_{2}\cos\theta_{n^{\prime}}-\sin a_{2}\sin
\theta_{n^{\prime}}\right]  \right\rangle _{\theta_{n},\theta_{n^{\prime}}%
}\nonumber\\
&  =\cos a_{1}\cos a_{2}\left\langle \cos\left[  \theta_{n}\right]
\cos\left[  \theta_{n^{\prime}}\right]  \right\rangle _{\theta_{n}%
,\theta_{n^{\prime}}}+\sin a_{1}\sin a_{2}\left\langle \sin\left[  \theta
_{n}\right]  \sin\left[  \theta_{n^{\prime}}\right]  \right\rangle
_{\theta_{n},\theta_{n^{\prime}}}\nonumber\\
&  -\sin a_{1}\cos a_{2}\left\langle \sin\left[  \theta_{n}\right]
\cos\left[  \theta_{n^{\prime}}\right]  \right\rangle _{\theta_{n}%
,\theta_{n^{\prime}}}-\cos a_{1}\sin a_{2}\left\langle \cos\left[  \theta
_{n}\right]  \sin\left[  \theta_{n^{\prime}}\right]  \right\rangle
_{\theta_{n},\theta_{n^{\prime}}}\nonumber\\
&  =\left(  1/2\right)  \delta_{nn^{\prime}}\left[  \cos a_{1}\cos a_{2}+\sin
a_{1}\sin a_{2}\right]  =\left(  1/2\right)  \delta_{nn^{\prime}}\cos\left(
a_{1}-a_{2}\right)  ,
\end{align}
from%
\begin{equation}
\left\langle \cos\left[  \theta_{n}\right]  \cos\left[  \theta_{n^{\prime}%
}\right]  \right\rangle =\left\langle \sin\left[  \theta_{n}\right]
\sin\left[  \theta_{n^{\prime}}\right]  \right\rangle =\left(  1/2\right)
\delta_{nn^{\prime}} \label{coscos}%
\end{equation}
\qquad and
\begin{equation}
\left\langle \cos\left[  \theta_{n}\right]  \sin\left[  \theta_{n^{\prime}%
}\right]  \right\rangle =0. \label{cossin}%
\end{equation}
We then summed over $n^{\prime}$ in%
\begin{align}
&  \left\langle \phi\left(  t_{1},x_{1}\right)  \phi\left(  t_{2}%
,x_{2}\right)  \right\rangle \nonumber\\
&  =\sum\nolimits_{n=-\infty}^{n=\infty}\sum\nolimits_{n^{\prime}=-\infty
}^{n^{\prime}=\infty}A_{n}A_{n^{\prime}}\left(  1/2\right)  \delta
_{nn^{\prime}}\cos\left[  k_{n}x_{1}-k_{n^{\prime}}x_{2}-\omega_{n}%
t_{1}+\omega_{n^{\prime}}t_{2}\right] \nonumber\\
&  =\frac{1}{2}\sum\nolimits_{n=-\infty}^{n=\infty}\,A_{n}^{2}\cos\left[
k_{n}\left(  x_{1}-x_{2}\right)  -\omega_{n}\left(  t_{1}-t_{2}\right)
\right]  .
\end{align}
\ 

\subsection{Lorentz-Invariant Random Radiation: Zero-Point Radiation}

Relativistic massless waves have the same speed $c$ in every inertial frame,
but have different frequencies and wavelengths, depending on the relativistic
inertial frame. \ Because waves allow superposition, we can ask whether there
is any spectrum of random classical waves with the same correlation function
in every inertial frame. \ The correlation function is a well-defined quantity
despite the randomness of the radiation. \ A distribution of random waves
which had the same correlation function in every inertial frame would be a
possible \textit{fundamental }homogeneous (source-free) solution for Maxwell's
equations. \ Such a fundamental homogeneous solution would affect \textit{all}
systems, including thermal radiation. Indeed, in the limit of an infinitely
large box, $a\rightarrow\infty$,\ where the sum in Eq. (\ref{pxtpxt}) becomes
an integral, there is a unique such spectrum (up to a multiplicative
constant). \ The correlation function for scalar classical zero-point
radiation has a spectral function $\left(  b\omega\right)  ^{-1}$ giving
\begin{equation}
\left\langle \phi_{zp}\left(  t_{1},x_{1}\right)  \phi_{zp}\left(  t_{2}%
,x_{2}\right)  \right\rangle =\frac{1}{2}\int_{-\infty}^{\infty}\frac
{dk\,}{b\omega}\cos\left[  k\left(  x_{1}-x_{2}\right)  -\omega\left(
t_{1}-t_{2}\right)  \right]  \label{zpup}%
\end{equation}
where $\omega=c\left\vert k\right\vert $. \ Under Lorentz transformation, this
becomes \
\begin{align}
\left\langle \phi_{zp}^{\prime}\left(  t_{1}^{\prime},x_{1}^{\prime}\right)
\phi_{zp}^{\prime}\left(  t_{2}^{\prime},x_{2}^{\prime}\right)  \right\rangle
&  =\frac{1}{2}\int_{-\infty}^{\infty}\frac{dk\,}{b\omega}\cos\left[
k^{\prime}\left(  x_{1}^{\prime}-x_{2}^{\prime}\right)  -\omega^{\prime
}\left(  t_{1}^{\prime}-t_{2}^{\prime}\right)  \right] \nonumber\\
&  =\frac{1}{2}\int_{-\infty}^{\infty}\frac{dk^{\prime}\,}{b\omega^{\prime}%
}\cos\left[  k^{\prime}\left(  x_{1}^{\prime}-x_{2}^{\prime}\right)
-\omega^{\prime}\left(  t_{1}^{\prime}-t_{2}^{\prime}\right)  \right]  ,
\label{zpp}%
\end{align}
since the phase is invariant under Lorentz transformation as in Eq.
(\ref{pha}), and since $dk/\omega=dk^{\prime}/\omega^{\prime}$ as follows from
Eqs. (\ref{omp}) and (\ref{kp}). \ But then the zero-point correlation
function for the random radiation takes exactly the same form in the unprimed
inertial frame (\ref{zpup}) and in the primed inertial frame (\ref{zpp}).
\ The spectrum here involving $b\omega$ is Lorentz invariant for any choice of
the constant $b$. \ 

\subsection{Two Fundamental Constants: $c$ and $\hbar$}

The specific spectrum of zero-point radiation gives a state of random
radiation which takes the same form in every inertial frame. \ Since the
random radiation is constantly fluctuating because of the interference between
waves of different frequencies, only the correlation function (and hence the
spectrum) are constant in space and in time.\ Here the spectrum corresponds to
$\left(  b\omega\right)  ^{-1}$ where $b$ is a \ constant. \ Thus, in the
presence of zero-point radiation, our classical theory of massless waves
involves two fundamental constants. \ The speed of light $c$ is associated
with the relativistic aspect of the theory. \ The second constant sets the
scale of the zero-point radiation, the scale of the special Lorentz-invariant
spectrum of \ source-free random radiation which has the same correlation
function in every relativistic inertial frame. \ The constant $b$ has
dimensions corresponding to the product of energy and time. \ The
determination of the constant $b$ is an experimental question in physics. \ 

Classical electromagnetic zero-point radiation leads to Casimir forces between
uncharged conducting parallel plates.\cite{Casimir} \ The Lorentz-invariant
functional dependence $\left\langle U_{n}\right\rangle =const\times\omega$ per
normal mode of the classical electromagnetic zero-point radiation gives the
correct functional form $1/d^{4}$ for the distance-dependence of the forces
between the plates at separation $d$ for low temperatures or small
separations. \ In order to have the classical theory give the entirely correct
numerical value for the force between the plates, we must choose the scale
constant as corresponding to an average energy per normal mode given by
$\left\langle U_{n}^{zp}\right\rangle =\left(  \hbar/2\right)  \omega$ where
$\hbar$ takes the same \textit{numerical} value as Planck's constant. \ We
will take this same energy-frequency connection for the zero-point radiation
for all relativistic waves. \ 

We note that in \textit{classical} physics, the symbol $\hbar$ is simply a
number setting the scale for \textit{relativistic waves} in the zero-point
spectrum. \ This \textit{classical} interpretation is completely different
from the interpretation given for $\hbar$ in \textit{quantum} theory where the
constant sets the scale of energy quanta. \ 

\subsection{Fluctuations in Random Radiation}

The interference of waves of differing frequencies gives rise to fluctuations
in total wave amplitude. \ When a large number of waves of closely-spaced
frequency interfere, the wave amplitude fluctuates in space and time, and so
leads to fluctuations in energy. \ If the collection of interfering waves is
analyzed by comparing the behavior in many long-but-finite time intervals of
equal length, the wave motion can be described in terms of the stochastic
behavior of the amplitude of the normal modes of oscillation.\cite{EH1910a}%
\cite{L1916}\cite{T1962}\cite{Rice} \ These references show that the wave
amplitudes of the normal modes satisfy a Gaussian distribution with average
value zero. \ Such Gaussian behavior can be shown to follow from the uniform
probability distribution for the random phases $\theta_{n}$ in Eq.
(\ref{phisum}) for waves which are closely-spaced in frequency. \ 

For \textit{individual} normal modes of frequency $\omega_{n}$ taken over a
finite-sized length $L$, the stochastic behavior is familiar from the
Boltzmann classical statistical mechanics of the harmonic oscillator where the
position $x$ and linear momentum $p$ have the (multivariable) Gaussian
probability distribution on phase space
\begin{equation}
P\left(  x,p\right)  dxdp=const\times\exp\left[  -\frac{p^{2}}{2mk_{B}%
T}\right]  \exp\left[  -\frac{m\omega^{2}x^{2}}{2k_{B}T}\right]  dxdp
\end{equation}
where $m$ is the mass of the oscillator of frequency $\omega$, $k_{B}$ is
Boltzmann's constant, and $T$ is the absolute temperature. \ We recall that
each \textit{radiation} normal mode has a Hamiltonian analogous to that a
harmonic oscillator. \ Accordingly, the energy distribution of a normal mode
is analogous to that of a Boltzmann harmonic oscillator, and has the
probability distribution
\begin{equation}
P\left(  U_{n},\overline{U_{n}}\right)  dU_{n}=\frac{1}{\overline{U_{n}}}%
\exp\left[  -\frac{U_{n}}{\overline{U_{n}}}\right]  dU_{n} \label{ProbUn}%
\end{equation}
where $\overline{U_{n}}$ is the average energy of the normal mode. \ 

The stochastic behavior of the wave normal mode labeled by the integer $n$
leads to fluctuations in the mode energy $\Delta U_{n}$ which are related back
to the average energy $\overline{U_{n}}$ of the normal mode as%
\begin{equation}
\left(  \Delta U_{n}\right)  ^{2}=\overline{\left[  U_{n}-\overline{U_{n}%
}\right]  ^{2}}=\overline{U_{n}^{2}-2U_{n}\overline{U_{n}}+\overline{U_{n}%
}^{2}}=\overline{U_{n}^{2}}-\overline{U_{n}}^{2}=\overline{U_{n}}^{2},
\label{DU2}%
\end{equation}
where in the last step, we have used the probability distribution in Eq.
(\ref{ProbUn}) to find%
\begin{equation}
\overline{U_{n}^{2}}=2\overline{U_{n}}^{2}.
\end{equation}
For \ zero-point radiation where the average energy is $\overline{U_{n}}%
=\hbar\omega_{n}/2$, this probability distribution for the energy per normal
mode takes the form%
\begin{equation}
P_{zp}(U_{n},\hbar\omega_{n}/2)dU_{n}=\frac{1}{\hbar\omega_{n}/2}\exp\left[
-\frac{U_{n}}{\hbar\omega_{n}/2}\right]  dU_{n}. \label{Pzp}%
\end{equation}

For each \textit{individual} radiation mode, the random behavior is correctly
given by the Boltzmann analysis as in Eq. (\ref{ProbUn}). \ However, the
average energy for modes of different frequency is an entirely different
matter. \ The average energy per normal mode for Lorentz-invariant zero-point
radiation was found to be $\overline{U_{n}}=\hbar\omega/2$ from an analysis
for \textit{relativistic waves} and the a comparison with Casimir forces at
low temperatures. \ Classical zero-point radiation is the most random
distribution of radiation within a relativistic theory. The zero-point
radiation spectrum has no preferred inertial frame, no preferred length, no
preferred time, no preferred energy. The one preferred quantity is the scale
$\hbar$ of the zero-point radiation, which is measured in erg-seconds. Next,
we wish to determine the average energy per normal mode for \textit{thermal}
radiation, which is \textit{not} relativistically invariant. \ \ 

\section{Thermal Radiation in Classical Theory}

\subsection{Preferred Inertial Frame of Thermal Equilibrium}

Classical \textit{electromagnetic zero-point radiation} takes the same form
(has the same correlation function) in \textit{every} inertial frame.
\ Therefore no inertial frame is distinguished by its zero-point radiation.
\ On the other hand, the equilibrium \textit{thermal radiation }of classical
physics has a preferred inertial frame, that of the container in which it is
in equilibrium. \ Thus zero-point radiation and thermal radiation are
completely different in their inertial-frame dependence. \ 

For our one-dimensional relativistic waves, we will assume that the thermal
equilibrium is analyzed over a finite length from $x=0$ and $x=L$. \ Our waves
may be analyzed in terms of a Fourier sine-series in space, and can be
expressed as a sum over the standing normal modes,
\begin{equation}
\phi(t,x)=\sum\nolimits_{n=1}^{\infty}A_{n}\sin\left(  \frac{n\pi x}%
{L}\right)  \cos\left[  \frac{n\pi ct}{L}-\theta_{n}\right]  .
\end{equation}
The energy of the waves now involves the same expression as in Eq. (\ref{HH}),
but the integral is from $x=0$ to $x=L$. \ The zero-point spectrum with
average wave energy $\overline{U_{zp}}=\hbar\omega/2$ requires an average
magnitude for the wave amplitude $\overline{\left\vert A_{n}^{zp}\right\vert
}=\sqrt{2\hbar c/\left(  n\pi\right)  },$
\begin{equation}
\phi_{zp}\left(  t,x\right)  =\sum\nolimits_{n=1}^{\infty}\sqrt{\frac{2\hbar
c}{n\pi}}\sin\left(  \frac{n\pi x}{L}\right)  \cos\left[  \frac{n\pi ct}%
{L}-\theta_{n}\right]  . \label{phizptx}%
\end{equation}
Thus, the amplitude of the relativistic wave in the length $L$ at temperature
$T=0$ does not vanish but has the fluctuation of a Gaussian random process
with an average (zero-point) energy $\hbar\omega/2$. \ The same average energy
per normal mode $\hbar\omega/2$ holds independent of the length $L$ of the
interval chosen.

\subsection{Fluctuations of Zero-Point Radiation}

As seen in Eq. (\ref{phizptx}), there are an infinite number of
zero-point-radiation normal modes, each with average energy $\hbar\omega
_{n}/2$. \ Thus, although the energy for each normal mode is finite and the
energy in any finite frequency interval \ from $n_{1}\pi c/L$ to $n_{2}\pi
c/L$ is finite, the total zero-point energy of the massless waves in the box
of length $L$ is divergent because there are an infinite number of radiation
modes. \ Furthermore, zero-point radiation is random radiation, and so each
normal mode has the fluctuations associated with the probability distribution
given in Eq. (\ref{Pzp}). \ However, in order to account for
experimentally-observed Casimir forces, the Lorentz-invariant zero-point
radiation is something which is \textit{always} present in a relativistic wave
system. \ The fluctuations of zero-point radiation are temperature-independent
fluctuations which are the same in every inertial frame and can not correspond
to the fluctuations of thermodynamic entropy.

\subsection{Fluctuations of Thermal Radiation}

Waves from different sources in the same region of spacetime allow linear
superposition. \ Thus the presence of both zero-point radiation $\phi
^{zp}\left(  t,x\right)  $ and additional thermal radiation $\phi^{T}\left(
t,x\right)  $ in a region of length $L$ will lead to a wave which is a linear
superposition\cite{Goldman65}%
\begin{equation}
\phi^{total}\left(  t,x\right)  =\phi^{T}\left(  t,x\right)  +\phi^{zp}\left(
t,x\right)  . \label{linwave}%
\end{equation}
Although the total wave amplitude involves linear addition of the
temperature-dependent thermal radiation and the temperature-independent
zero-point radiation, the total energy is quadratic\cite{Goldman65q} in the
total wave amplitude
\begin{equation}
\left[  \phi^{total}\left(  t,x\right)  \right]  ^{2}=\left[  \phi^{T}\left(
t,x\right)  \right]  ^{2}+2\phi^{T}\left(  t,x\right)  \phi^{zp}\left(
t,x\right)  +\left[  \phi^{zp}\left(  t,x\right)  \right]  ^{2}.
\end{equation}
The square in the wave amplitude is directly related to the energy. \ The
amplitudes are assumed Gaussian with vanishing average value. \ If the total
radiation is regarded as random radiation arising from two
\textit{independent} waves, then, when averaged, the cross-term contribution
$\phi^{T}\left(  t,x\right)  \phi^{zp}\left(  t,x\right)  $ is the product of
the amplitudes from the independent sources and so vanishes, $\left\langle
\phi^{T}\left(  t,x\right)  \phi^{zp}\left(  t,x\right)  \right\rangle
=\left\langle \phi^{T}\left(  t,x\right)  \right\rangle \left\langle \phi
^{zp}\left(  t,x\right)  \right\rangle =0$. \ Thus we have the average total
energy as simply the sum of the average energies,%
\begin{equation}
\overline{U^{total}}=\overline{U^{T}}+\overline{U^{zp}} \label{Utotalav}%
\end{equation}
with no interference evident in the average energies.

However, the \textit{fluctuations} in the \textit{energy} will involve the
fourth power of the total field $\phi^{total}\left(  t,x\right)  $, and thus
will include a cross-terms of the form $\left[  \phi^{T}\left(  t,x\right)
\phi^{zp}\left(  t,x\right)  \right]  ^{2}=\left[  \phi^{T}\left(  t,x\right)
\right]  ^{2}\left[  \phi^{zp}\left(  t,x\right)  \right]  ^{2}$. \ When
averaged, this cross-term is nonvanishing, $\left\langle \left[  \phi
^{T}\left(  t,x\right)  \phi^{zp}\left(  t,x\right)  \right]  ^{2}%
\right\rangle =\left\langle \left[  \phi^{T}\left(  t,x\right)  \right]
^{2}\right\rangle \left\langle \left[  \phi^{zp}\left(  t,x\right)  \right]
^{2}\right\rangle \neq0,$ corresponding (for independent $\phi^{T}$ and
$\phi^{zp}$ fluctuations) to
\begin{equation}
\overline{\left[  U^{total}\right]  ^{2}}=\overline{\left[  U^{T}%
+U^{zp}\right]  ^{2}}=\overline{\left[  U^{T}\right]  ^{2}}+2\overline{U^{T}%
}\,\overline{U^{zp}}+\overline{\left[  U^{zp}\right]  ^{2}}.
\end{equation}
Thus due to the interference cross-term, the fluctuation in the total energy
will be different from the fluctuation in energy in the absence of the other
wave. \ This interference between two independent sources of radiation is the
basis of [Hanbury-Brown and Twiss] interferometry (used in astronomy), which
can be explained purely classically, but is often interpreted in terms of
quantum photon bunching.\cite{Garg} \ 

\subsection{Thermodynamic Entropy and Fluctuations Above Zero-Point
Fluctuations\ }

In order to correspond to part of a thermodynamic system, a finite length $L$
containing random massless waves must have a finite amount of thermal energy
$U^{T}$ which will be associated with a non-zero temperature $T$ and a finite
thermodynamic entropy $S$. \ The total thermodynamic energy $U^{T}$ is spread
across the normal modes labeled by the integer $n$ so that each mode has a
thermodynamic energy $U_{n}^{T}(T)$ and associated thermodynamic entropy
$S_{n}\left(  T\right)  $. \ Both the total thermodynamic energy and the total
thermodynamic entropy are sums over the thermodynamic energy and thermodynamic
entropy of each normal mode. \ Since we are dealing with averages over
\textit{waves} from independent sources for each normal mode, the
thermodynamic energy $U_{n}^{T}(T)$ for the mode $n$\ will follow the same
rule involving waves as appears in Eq. (\ref{Utotalav}),
\begin{equation}
\overline{U_{n}^{total}(T)}=\overline{U_{n}^{T}(T)}+\overline{U_{n}^{zp}}.
\label{Ucalt}%
\end{equation}
The total average thermodynamic energy in the box is simply the sum of the
average thermodynamic energies in each mode. \ 

On the other hand, each normal mode will therefore have an magnitude of its
amplitude $\left\vert A_{n}\right\vert ,$ larger than the zero-point amplitude
$\left\vert A_{n}^{zp}\right\vert =\sqrt{2\hbar c/(n\pi)}$. \ But then from
Eqs. (\ref{DU2}) and (\ref{Ucalt}), the energy fluctuations of each normal
mode will be larger than those of the zero-point radiation with
\begin{align}
\left[  \Delta U_{n}^{total}(T)\right]  ^{2}  &  =\overline{U_{n}^{total}}%
^{2}=\overline{\left[  U_{n}^{T}(T)+U_{n}^{zp}\right]  }^{2}\nonumber\\
&  =\overline{U_{n}^{T}(T)}^{2}+2\overline{U_{n}^{T}(T)}\,\overline{U_{n}%
^{zp}}+\overline{U_{n}^{zp}}^{2}. \label{DUtot}%
\end{align}
The presence of the zero-point radiation does not appear in the average
energy, but it does appear in the \textit{fluctuations} of the thermodynamic energy.

\subsection{Connecting Entropy and Temperature}

\subsubsection{The Rayleigh-Jeans Spectrum in the Absence of Zero-Point
Radiation}

The entropy of the distribution of thermal radiation is associated with the
distribution of the thermodynamic energy $U^{T}$ among the normal modes so as
to produce a maximum entropy. \ For this to occur, the second derivative of
the thermodynamic entropy for each mode with respect to the thermodynamic
energy must be associated with fluctuations in the thermodynamic energy. \ The
traditional connection (when zero-point energy is \textit{not} considered)
is\cite{E1909a}\cite{B1969fl}\cite{Sciama}
\begin{equation}
\frac{\partial^{2}S_{n}}{\partial\overline{U_{n}^{T}}^{2}}=\frac{-k_{B}%
}{\Delta U_{n}^{T}}=\frac{-k_{B}}{\overline{U_{n}^{T}}^{2}}.\text{ \ No
zero-point radiation.} \label{Nzpr}%
\end{equation}
Upon integrating this equation once, we find
\begin{equation}
\frac{\partial S_{n}}{\partial\overline{U_{n}^{T}}}=\frac{k_{B}}%
{\overline{U_{n}^{T}}}.\text{ \ No zero-point radiation.}%
\end{equation}
Then using $\partial S_{n}/\partial\overline{U_{n}^{T}}=1/T\,$, we immediately
obtain the equipartition energy expression
\begin{equation}
\overline{U_{n}^{T}}=k_{B}T.\text{ \ No zero-point radiation.}%
\end{equation}
\ This result gives the Rayleigh-Jeans spectrum. \ The Rayleigh-Jeans result
shows no basis for assigning more thermal energy to low frequencies and less
to high frequencies. \ All frequencies are treated in the same manner, and the
result is a divergent total thermal energy, \textquotedblleft the ultraviolet
catastrophe.\textquotedblright

\subsubsection{The Planck Spectrum in the Presence of Zero-Point Radiation}

However, if temperature-independent zero-point radiation is indeed present,
the expression in Eq. (\ref{Nzpr}) cannot be correct. \ The energy
fluctuations will be larger corresponding to both zero-point energy and the
thermodynamic energy appearing as superimposed waves as in Eq. (\ref{linwave}%
). \ Now, the zero-point radiation is a temperature-independent source of
noise for the thermodynamic system. \ Using Eq. (\ref{DUtot}) and following
the ideas for wave information in the presence of random noise, it is natural
to associate the thermodynamic entropy $S_{n}$ with only the energy
fluctuations above the zero-point energy fluctuations for each normal mode,
\begin{equation}
\left[  \Delta U_{n}^{T}\right]  ^{2}=\overline{U_{n}^{total}}^{2}%
-\overline{U_{n}^{zp}}^{2}=\overline{U_{n}^{T}(T)}^{2}+2\overline{U_{n}%
^{T}(T)}\,\overline{U_{n}^{zp}}%
\end{equation}
since the zero-point radiation is the same in every inertial frame and has no
thermodynamic entropy of its own. \ Using this expression, we see that the
thermodynamic fluctuations vanish when there is no thermal energy,
$\overline{U_{n}^{T}}=0$, and go toward $\overline{U_{n}^{T}}^{2}$ when
$\overline{U_{n}^{T}}>>\overline{U_{n}^{zp}}$. \ Therefore we write instead of
Eq. (\ref{Nzpr})
\begin{equation}
\frac{\partial^{2}S_{n}}{\partial\overline{U_{n}^{T}}^{2}}=\frac{\partial
^{2}S_{n}}{\partial\overline{U_{n}^{total}}^{2}}=\frac{-k_{B}}{\overline
{U_{n}^{total}}^{2}-\overline{U_{n}^{zp}}^{2}}.
\end{equation}
Holding the (average) zero-point energy of the normal mode as fixed
(independent of the thermal energy), and integrating this equation once with
respect to the thermal energy gives%

\begin{equation}
\frac{\partial S_{n}}{\partial\overline{U_{n}^{T}}}=\frac{\partial S_{n}%
}{\partial\overline{U_{n}^{total}}}=\frac{k_{B}}{2\overline{U_{n}^{zp}}}%
\ln\left(  \frac{\overline{U_{n}^{total}}+\overline{U_{n}^{zp}}}%
{\overline{U_{n}^{total}}-\overline{U_{n}^{zp}}}\right)  .
\end{equation}
Thus, since $\partial S_{n}/\partial\overline{U_{n}^{T}}=1/T$, we have%
\begin{equation}
\frac{1}{T}=\frac{k_{B}}{2\overline{U_{n}^{zp}}}\ln\left(  \frac
{\overline{U_{n}^{total}}+\overline{U_{n}^{zp}}}{\overline{U_{n}^{total}%
}-\overline{U_{n}^{zp}}}\right)
\end{equation}
or, introducing $\overline{U_{n}^{zp}}=\hbar\omega/2$,
\begin{equation}
\frac{1}{T}=\frac{k_{B}}{\hbar\omega}\ln\left(  \frac{\overline{U_{n}^{total}%
}+\hbar\omega/2}{\overline{U_{n}^{total}}-\hbar\omega/2}\right)  .
\end{equation}
Solving for $\overline{U_{n}^{total}}$ gives%
\begin{equation}
\overline{U_{n}^{total}}=\frac{\hbar\omega}{2}\left\{  \frac{\exp\left[
\hbar\omega/\left(  k_{B}T\right)  \right]  +1}{\exp\left[  \hbar
\omega/\left(  k_{B}T\right)  \right]  -1}\right\}
\end{equation}
or
\begin{equation}
\overline{U_{n}^{total}}=\frac{\hbar\omega}{2}\coth\left(  \frac{\hbar\omega
}{2k_{B}T}\right)  =\frac{\hbar\omega}{\exp\left[  \hbar\omega/\left(
k_{B}T\right)  \right]  -1}+\frac{1}{2}\hbar\omega,
\end{equation}
which is exactly the Planck spectrum including zero-point radiation.
\ The\ \textit{Lorentz-invariant zero-point radiation} with its increasing
energy and larger fluctuations at higher frequencies provides the criterion
which distinguishes low and high frequencies. \ Including classical
electromagnetic zero-point radiation introduces the constant $\hbar$ and
provides a natural classical explanation for the Planck spectrum. \ 

\section{Comments}

\subsection{Two Versions of Classical Electrodynamics}

Classical electrodynamics appears in two different fundamental forms. \ The
usual textbook treatment assumes that there is no zero-point radiation
present. \ This version forms the basis for electromagnetic technology at the
macroscopic scale. \ The second version of classical electrodynamics includes
classical electromagnetic zero-point radiation as a fundamental
aspect.\cite{Bzp} \ As shown in the present article, the presence of
zero-point radiation within classical theory allows the theory to describe
some aspects of nature which are usually treated within quantum theory.
\ Thus, at all temperatures when classical zero-point radiation is included,
there are natural classical explanations for Casimir forces, van der Waals
forces, low-temperature decrease of specific heats of solids, blackbody
radiation, the absence of atomic collapse, and superfluid-like
behavior.\cite{SED} \ The presence of Planck's constant $\hbar$ comes from the
scale of zero-point radiation. \ There are no quanta in the classical theory.

Students in our courses have no idea that a \textit{Lorentz-invariant
spectrum} of classical radiation corresponds to a constant times frequency
$\omega$ per normal mode. \ Students do not realize that a radiation spectrum
with energy $\hbar\omega/2$ corresponds to a Lorentz-invariant spectrum of
classical radiation. \ Indeed, a charged classical harmonic oscillator in
equilibrium with classical zero-point radiation will have an average energy
$\hbar\omega/2$, corresponding to the average energy of the zero-point
radiation modes at the same frequency as the natural frequency of the
oscillator. \ 

On the other hand, quantum electrodynamics (which \textit{is} relativistic)
does include quantum zero-point radiation. \ However, the perspective on
zero-point radiation is quite different between classical and quantum
theories. \ \textit{Classical} electrodynamics regards Lorentz-invariant
classical zero-point radiation as the basis for introducing Planck's constant
$\hbar$ as a \textit{scale factor}, and so allowing descriptions of phenomena
involving this constant. \ \textit{Quantum} theory regards the idea of
\textit{quanta proportional to }$\hbar$ as fundamental, and zero-point
radiation or zero-point energy follow from the quantum ideas. \ 

Classical zero-point radiation does not share the \textquotedblleft intrinsic
randomness\textquotedblright\ of quantum measurement theory. \ Thus, if the
random phases of all the zero-point radiation waves were known, then the
evolution of the electromagnetic radiation field would be completely
determined. \ The situation is analogous to the statistical randomness
corresponding to the unknown positions and velocities of the molecules in a
classical gas, which provide the background for Boltzmann's thermodynamics. \ 

\subsection{Planck Spectrum within Classical Electrodynamics}

Although \textit{nonrelativistic} quantum theory includes the idea of
zero-point \textit{energy}, it does not include zero-point \textit{radiation}
because zero-point radiation requires the ideas of relativity. \ The Planck
spectrum developed from the quantum statistical mechanics of photons, does
\textit{not} include zero-point radiation. \ Indeed, the Planck spectrum given
by quantum photon ideas goes to zero at temperature $T=0$. \ 

Classical zero-point radiation is also needed for the completion of the Planck
spectrum within classical theory. \ If the Planck spectrum without zero-point
radiation is applied within classical electrodynamics, there is no temperature
at which the Casimir forces are given correctly.\cite{B1975c} \ Furthermore,
within classical theory, the Planck spectrum which omits the zero-point
contribution does not give the equipartition result at high temperature and
fixed frequency, since the zero-point contribution is subtracted off the
equipartition value as
\begin{equation}
\overline{U_{n}^{T}}=\frac{\hbar\omega_{n}}{\exp\left[  \hbar\omega
_{n}/\left(  kT\right)  \right]  -1}\approxeq kT-\frac{\hbar\omega_{n}}%
{2}+kT\left\{  O\left[  \left(  \frac{\hbar\omega_{n}}{kT}\right)
^{2}\right]  \right\}
\end{equation}
for $kT>\hbar\omega$. \ Thus, even at high temperatures, the Planck spectrum
\textit{without} zero-point radiation does not give the functional dependence
for Casimir forces corresponding to the Rayleigh-Jeans spectrum.\cite{B1975c}
\ The inclusion of zero-point radiation brings the average radiation energy
into agreement with energy equipartition for $kT>>\hbar\omega$. \ The
classical Planck spectrum \textit{including} zero-point radiation gives
consistent electromagnetic Casimir forces at all parallel-plate separations at
all\ temperatures,\cite{B1975c} in complete agreement with quantum
electrodynamic calculations. \ 

Current textbooks of modern physics do not mention special relativity in
connection with the Planck spectrum.\cite{mod} \ In fact, the textbooks often
apply the \textit{equipartition} theorem of \textit{nonrelativistic
particle-based waves} to relativistic electromagnetic waves so as to obtain
the Rayleigh-Jeans spectrum.\cite{mod}\cite{B2018c} Thus the crucial contrasts
between relativistic wave motion and nonrelativistic particle motion are not
suggested to students, and they have no basis for understanding the Planck
spectrum within the context of classical physics. \ \ 

\

\bigskip

SymmetryPlanck11.tex \ \ December 1, 2023

\end{document}